\documentclass[reprint, amsmath, amssymb, aps, pra, superscriptaddress]{revtex4-2}

\usepackage[colorlinks=true]{hyperref}
\usepackage{bm}
\usepackage{subcaption}
\usepackage{graphicx}
\usepackage{soul}
\usepackage{comment}
\usepackage{ulem}
\usepackage{xcolor}

\usepackage{caption}   % For captions

\usepackage{adjustbox}

\begin{document}

\preprint{APS/123-QED}

\title{Generating entangled pairs of vortex photons via induced emission}% Force line breaks with \\

\author{D.\,V.~Grosman}
\email{dmitriy.grosman@metalab.ifmo.ru}
\affiliation{School of Physics and Engineering,
ITMO University, 197101 St. Petersburg, Russia}

\author{G.\,K.~Sizykh}
\email{georgii.sizykh@metalab.ifmo.ru}
\affiliation{School of Physics and Engineering,
ITMO University, 197101 St. Petersburg, Russia}
\affiliation{Petersburg Nuclear Physics Institute of NRC “Kurchatov Institute”, 188301 Gatchina, Russia}

\author{E.\,O.~Lazarev}
\email{egor.lazarev@metalab.ifmo.ru}
\affiliation{School of Physics and Engineering,
ITMO University, 197101 St. Petersburg, Russia}

\author{G.\,V.~Voloshin}
\email{gavriilvsh@gmail.com}
\affiliation{Peter the Great St. Petersburg Polytechnic University, 195251 St. Petersburg, Russia}

\author{D.\,V.~Karlovets}
\email{dmitry.karlovets@metalab.ifmo.ru}
\affiliation{School of Physics and Engineering,
ITMO University, 197101 St. Petersburg, Russia}
\affiliation{Petersburg Nuclear Physics Institute of NRC “Kurchatov Institute”, Gatchina, Russia}

\begin{abstract}
Pairs of entangled vortex photons can promise new prospects of application in quantum computing and cryptography. We investigate the possibility of generating such states via two-level atom emission induced by a single photon wave packet with a definite total angular momentum (TAM). The entangled pair produced in this process possesses well-defined mean TAM with the TAM variation being much smaller than $\hbar$. On top of that, the variation exponentially decreases with the increase in TAM of the incident photon. Our model allows one to track the time evolution of the state of the entangled pair. An experimentally feasible scenario is assumed, in which the incident photon interacts with a spatially confined atomic target. We conclude that induced emission can be used as a source of entangled vortex photons with applications in atomic physics experiments, quantum optics, and quantum information sciences.
\end{abstract}

\maketitle

\section{Introduction}
Advances in singular optics enable producing and manipulating wave packets of photons with orbital angular momentum (OAM) \cite{Padgett2017May, Knyazev2018Feb, Franke-Arnold2008Aug}. These vortex photons have found applications in many areas such as microscopy, nanoparticle manipulation, biomechanics and etc.  \cite{Allen1999, Franke-Arnold2008Aug, Padgett2014, Padgett2017May}. In addition, optical vortices are rapidly gaining popularity in quantum information sciences. The main advantage of quantum key distribution schemes based on the OAM degree of freedom is the ability to transmit more information and increase error tolerance \cite{sit2018quantum}. 

To establish secure communication entanglement based protocols are commonly used \cite{bacco2020quantum, PhysRevLett.67.661}. Over the past two decades, remarkable progress has been made in generating OAM-entangled photon pairs. Early demonstrations based on spontaneous parametric down-conversion (SPDC) in nonlinear crystals established the feasibility of high-dimensional entanglement with up to hundreds of spatial modes \cite{dada2011experimental, krenn2014generation, malik2016multi, erhard2018experimental},  providing a foundation for quantum communication and information protocols. More recently, efforts  have turned toward integrated photonic architectures and on-chip vortex emitters \cite{zhao2024integrated, Wang2024Aug}, which offer compactness, scalability, and rapid reconfigurability. At the same time, these approaches typically have low pair generation rate, are often broadband, and provide only limited control over the precise mode composition. In contrast, the atom-based induced-emission mechanism discussed here offers a  complementary route: it naturally yields narrowband biphoton states with well-defined and tunable TAM,  directly linked to the properties of the incident photon and the atomic transition. The generation probability in this approach can be controlled by the size of the atomic localization region.

\begin{figure}[h!]
\begin{subfigure}{0.45\textwidth}
\includegraphics[width=1\linewidth]{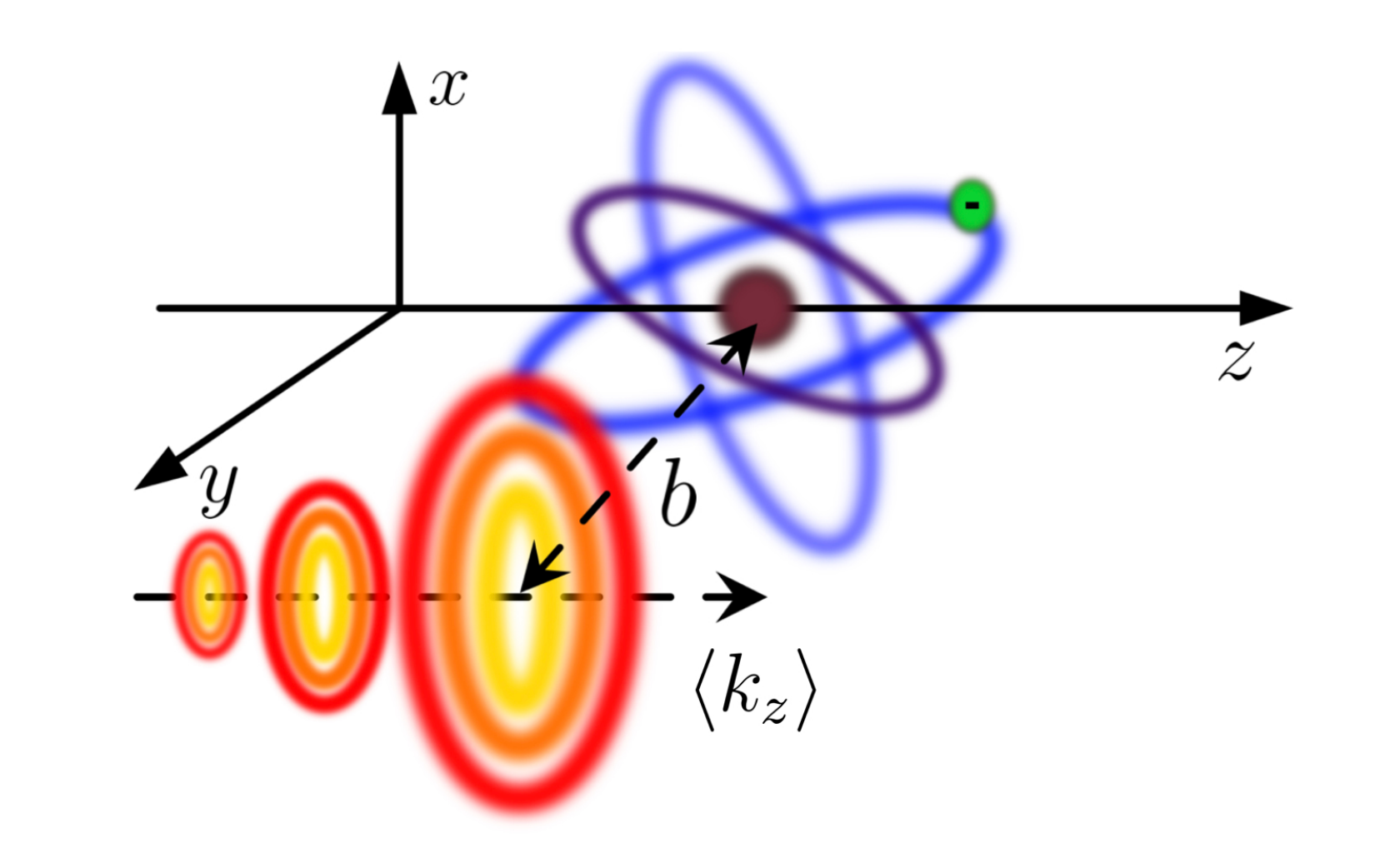}
\caption{}
\label{fig:enter-label_a}
\end{subfigure}
\begin{subfigure}{0.45\textwidth}
\includegraphics[width=0.9\linewidth]{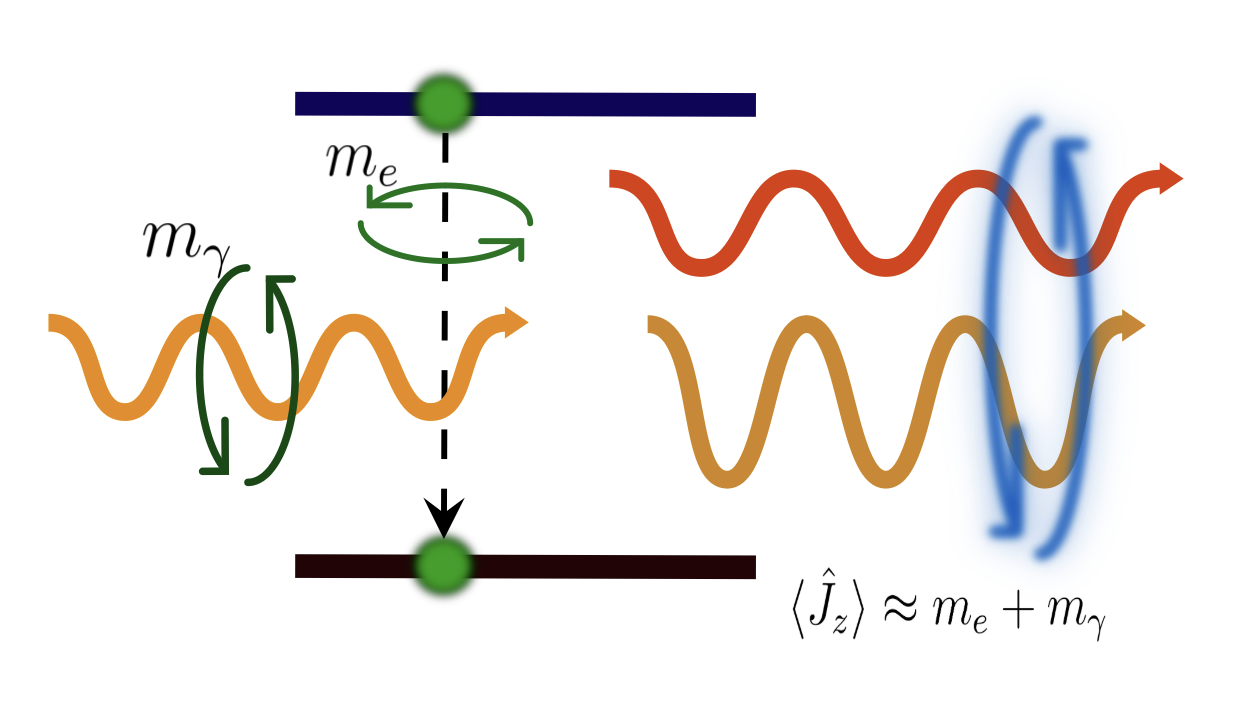}
\caption{}
\end{subfigure}
    \caption{Schematic of induced emission. (a) An incident photon wave packet $|\gamma_i\rangle$ with a mean longitudinal momentum $\langle k_z \rangle$ propagates along the $z$-axis at a distance $b$ from the center of mass of the atom. (b) Interaction of the incident photon with an excited atom triggers emission of an entangled pair of photons $|\gamma_1;\gamma_2\rangle$. The rings indicate the angular momentum of the initial photon, the electron in the excited state and the angular momentum of the final two photon state. For the two photon state blur is used to indicate that an offset between the incident photon and the atom result in a smear of the angular momentum around the mean value that is close to $m_\gamma+m_e$}.
    \label{fig:enter-label}
\end{figure}

A simple process giving rise to a pair of entangled photons involves an excited atom interacting with an incident single-photon electromagnetic field. The quantum nature of the process ensures that the emitted and incident photons share correlated properties such as polarization and angular momentum. A common way to describe the evolution of such a system is to either use the perturbation theory \cite{Scully1997Sep, Mandel1995Sep, Chang1971Aug, Dalibard1983Dec, Yeh1981Jun, Elyutin2012Mar} or consider a highly-populated mode of the incident electromagnetic field \cite{Stroud1971Mar, Smithers1975Dec}. Among other approaches is a Weisskopf-Wigner like approximation, developed in \cite{Bergmann1967Feb}, which is similar to the original calculations of Weisskopf and Wigner \cite{Weisskopf1930_1, Weisskopf1930_2}.

The interaction of vortex light with a two-level atom was considered by several authors with a classical description of electromagnetic field \cite{Picon2010Aug, Picon2010Feb, Jauregui2015Apr, Rodrigues2016Mar} and in a fully quantum \cite{Scholz-Marggraf2014Jul, Afanasev2013Sep, Afanasev2014Jan, Afanasev2016Jun, Peshkov2016May, Baragiola2012Jul, Wang2011Jun, Stobinska2009Apr} picture. Namely, Picon and colleagues investigated the state of the electron after the atomic ionization by a laser pulse \cite{Picon2010Aug, Picon2010Feb}. In works \cite{Jauregui2015Apr, Rodrigues2016Mar} the authors analyzed the influence of the laser beam shape on atomic transition rates. An interaction of a time-dependent Gaussian photon packet with a two-level system was considered in \cite{Baragiola2012Jul, Wang2011Jun, Stobinska2009Apr}. In \cite{Scholz-Marggraf2014Jul, Afanasev2013Sep, Afanasev2014Jan, Afanasev2016Jun, Peshkov2016May} photoexcitation and photoionization cross sections by vortex photons were derived and in \cite{Pavlov2024Sep} the preselected state of the photoelectron was obtained. One of the features common for all these studies is the change of the selection rules. The OAM of photons, in addition to their polarization, permits higher multipole transitions in atoms.

In this work we study interaction of an excited two-level atom with an incident vortex photon resulting in an entangled two-photon state. The schematics of the processes is depicted in Fig. \ref{fig:enter-label}. The incident photon is described by a state vector 

Our study focuses on the transfer of TAM from the incident photon to the entangled pair. We analyze the dependence of the mean value $\langle \hat{J}_z \rangle$ and the standard deviation of TAM $\sqrt{\langle \hat{J}^2_z\rangle - \langle \hat{J}_z\rangle^2}$ on experimental parameters such as the coherence length of the incident photon and the localization of the atomic target. To do so, we derive the state of the whole system, consisting of the atom and quantized electromagnetic field, at an arbitrary time. We solve the time-dependent Schr\"{o}dinger equation with initial conditions corresponding to the atom in the excited state and a Bessel-Gaussian vortex photon. Our results indicate that entangled photons can be generated via induced emission providing a technique alternative to the commonly used spontaneous parametric down-conversion.

\section{Theoretical framework}
\label{Sec:Theory}
\subsection{Interaction of the atom with the incident photon}
\label{subsec21}

The interaction of a two-level atom with quantized electromagnetic field is commonly described in the interaction picture \cite{Schleich2001Feb}
\begin{equation}
\label{int}
    i\partial_{t} | \psi(t) \rangle =  \sum\limits_{\nu,n}\left[ g_{\nu,n}^* \hat{\sigma}_{+,n}\hat{a}_{\nu}e^{i\Delta_{\nu}t} + \text{h.c.} \right]| \psi(t) \rangle.
\end{equation}
We assume the ground state $|\text{g} \rangle$ of the atom to be an $s-$orbital and for the excited state we take into account the magnetic sublevels $|\text{e}_{n}\rangle$, labeled by $n$. In Eq. \eqref{int} rotating wave approximation (RWA) is implied and $\nu = \{\bm{k}_{\nu},\lambda_{\nu}\}$ is the multi-index containing photon momentum $\bm{k}_{\nu}$ and polarization $\lambda_{\nu}$, $\hat{\sigma}_{+,n} = |\text{e}_n\rangle\langle \text{g}|$, $\hat{a}_{\nu}$ is the photon annihilation operator, $\Delta_{\nu} = \omega_{\nu} - \omega_{\text{a}}$ is the frequency detuning of the mode $\nu$ from the atomic resonance $\omega_{\text{a}}$.

In quantum optics the dipole approximation is also often assumed when describing the interaction of light with matter. However, within the dipole approximation occurs only between the incident photon and the atom center of mass \cite{PhysRevLett.89.143601}. 
The quadrupole transition is the lowest-order electric multipole process that involves the internal motion of the electron in OAM transfer. Therefore, we consider the exact non-dipolar coupling between electromagnetic field and the atom,
\begin{equation}
\label{coup}
    g_{\nu,n} = -\frac{e}{m}\sqrt{\frac{2\pi}{V\omega_{\nu}}}\langle \text{g}| \bm{e}_{\bm{k},s}\cdot\hat{\bm{p}}e^{i\bm{k}\cdot\bm{r}}| \text{e}_{n} \rangle,
\end{equation}
that accounts for \textit{all the multipole processes}. Here $V$ is the normalization volume. We emphasize, that Eq. \eqref{coup} is suitable for describing transitions between any two atomic states, regardless of their OAM. In the dipole approximation, on the other hand, one can only describe transitions between states with OAM differing by no more than unity.

Since the interaction Hamiltonian \eqref{int} changes the number of quanta of light by one, the natural anzats for the solution of the Schr\"{o}dinger equation is
\begin{equation}
\label{st}
\begin{aligned}
    |\psi(t) \rangle = \sum\limits_{\nu,n}&C_{e,\nu,n}(t)|\text{e}_{n}\rangle|\nu\rangle \\
    & + \sum\limits_{\nu_1,\nu_2}C_{g,\nu_1,\nu_2}(t)|\text{g}\rangle|\nu_1;\nu_2\rangle.
\end{aligned}
\end{equation}
The coefficient $C_{e,\nu,n}(t)$ is the probability amplitude of finding the atom in the excited state with the $z$-projection of OAM $n$ and a photon in the state $|\nu\rangle$. In turn, $C_{g,\nu_1,\nu_2}(t)$ corresponds to the probability amplitude of finding the atom in the ground state and two photons $|\nu_1;\nu_2\rangle \equiv 1/\sqrt{2}\left[|\nu_1\rangle|\nu_2\rangle + |\nu_2\rangle|\nu_1\rangle\right]$. The Schr\"{o}dinger equation and ansatz \eqref{st} lead to the following equation: 
\begin{multline}
\label{diffuclt equation}
    \dot{C}_{e,\nu,n}(t) = - \sum\limits_{\nu',n'}\int\limits_{0}^{t}d\tau\big[C_{e,\nu,n'}(\tau)g_{\nu',n'}e^{-i\Delta_{\nu'}t} + \\
    C_{e,\nu',n'}(t)g_{\nu,n'}e^{-i\Delta_{\nu}t} \big] g_{\nu',n}^* e^{i\Delta_{\nu'}t}.
\end{multline}
The expression for $C_{g,\nu_1,\nu_2}(t)$ is uniquely determined by $C_{e,\nu,n}(t)$ (see Sec. I of the Supplemental Material \cite{supp}).

We note that the set of possible photons modes in Eq.~\eqref{st} is assumed to be continuous as the problem is considered in free space rather than inside the resonator. In a cavity, if a photon is emitted in the same mode as the incident photon, the transition matrix element is enhanced by a factor of $\sqrt{N_{\text{ph}}}$, where $N_{\text{ph}}$ is the number of photons in the mode. This enhancement serves as the fundamental principle underlying laser operation. However, it is essential to recognize that the increase in the population of the incident photon mode is exclusively a characteristic of stimulated emission within a cavity. In free space, the contribution from two photons being emitted in the same mode is negligibly small, as the probability of this process is proportional to the density of final states (see Eqs. (33), (34) in \cite{Elyutin2012Mar} and the text around for details). Therefore, throughout this paper the term \textit{induced emission} refers to the processes in which an excited atom emits a photon in the presence of a given initial photon and should not be understood as stimulated emission in a cavity leading to mode amplification.

The unique solution to the Schr\"{o}dinger equation is determined by a specific initial condition. Suppose the atom is confined in a magnetic trap, and the photon, prepared in the state $|\gamma_i\rangle $,  propagates towards the trap. A pump field creates a population inversion in the atom and when the photon reaches it, the electron is in the excited state with OAM $m_e$. In such a scenario the state of the system at the time $t = 0$ is disentangled:
\begin{equation}
\label{initial conditions}
    |\psi(0) \rangle = |\text{e}_{m_e}\rangle|\gamma_{i}\rangle.
\end{equation}
This initial condition determines the influence of the incident photon characteristics on the state of the system and eventually on the state of the entangled pair of photons.

To solve Eq. \eqref{int} with initial condition \eqref{initial conditions}, it is convenient to decompose the state of the incident photon into the plane-wave basis: $|\gamma_i\rangle = \sum\limits_{\nu_0}\gamma(\nu_0)|\nu_0\rangle$. Then, the linearity of the Schr\"{o}dinger equation enables us to represent the state of the system at arbitrary time as
\begin{equation}
\begin{aligned}
    |\psi(t) \rangle = \sum\limits_{\nu_0}\gamma(\nu_0) |\psi^{\text{pw}}(t)\rangle&,\\
|\psi^{\text{pw}}(t) \rangle = \sum\limits_{\nu,n}C^{\text{pw}}_{e,\nu,n}(t)&|\text{e}_n\rangle|\nu\rangle \\
     + \sum\limits_{\nu_1,\nu_2}C^{\text{pw}}_{g,\nu_1,\nu_2}&(t)|\text{g}\rangle|\nu_1;\nu_2\rangle.
\end{aligned}
\end{equation}
Here, $|\psi^{\text{pw}}(t) \rangle $ is the state of the system for a plane wave incident photon and the probability amplitude $C_{e,\nu,n}^{\text{pw}}(t)$ is the solution to Eq. \eqref{diffuclt equation}. At $t = 0$ the state of the system is $|\psi^{\text{pw}} (0) \rangle = |\text{e}_{m_e}\rangle|\nu_0\rangle$, which is equivalent to $C_{e,\nu,n}^{\text{pw}}(0) = \delta_{n,m_e}\delta_{\nu,\nu_0}$, $C_{g,\nu_1, \nu_2}^{\text{pw}}(0) = 0$, and the amplitude $C_{e,\nu,n}^{\text{pw}}(t)$ reads (see Sec. II of the Supplemental Material \cite{supp} for details)
\begin{equation}
\label{Ce}
\begin{aligned}
    C^{\text{pw}}_{e,\nu,n}(t) & = \delta_{\nu,\nu_0}\delta_{n,m_e}e^{-\frac{\Gamma}{2}t} - g_{\nu,m_e}g^{*}_{\nu_0,n} e^{-\frac{\Gamma}{2}t}\times\\
    &\int\limits_{0}^{t}dt_2 \int\limits_{0}^{t_2}e^{\frac{\Gamma}{2}t_2 + i\Delta_{\nu_0}t_2 - \frac{\Gamma}{2}t_1 - i\Delta_{\nu}t_1} d t_1.
\end{aligned}
\end{equation}
Here, $\Gamma$ is the spontaneous decay rate of the excited atom and is typically much smaller than the frequency of the atomic resonance, $\Gamma \ll \omega_{\text{a}}$. The ground state amplitude $C_{g,\nu_1,\nu_2}^{\text{pw}}(t)$ can be found directly from $C_{e,\nu,n}^{\text{pw}}(t)$ (see Sec. I of the Supplemental Material \cite{supp}).

\subsection{Vortex incident photon}
\label{subsec22}

Now let us discuss an incident photon approaching the atom at an impact parameter $\bm{b} = \{b\cos\varphi_b, b\sin\varphi_b \}$ (see Fig. \ref{fig:enter-label_a}). We describe it as a localized Bessel - Gaussian wave packet with the wave function in momentum space
\begin{equation}
\label{nu0}
\begin{aligned}
    \langle \nu_0 |\gamma_i \rangle = N&\int 
    e^{-(\kappa - \kappa_\text{c})^2/2\sigma_{\kappa}^2}
    e^{ - (k_z - k_{\text{c}})^2/2\sigma_z^2}\\
    &e^{i\kappa b \cos(\varphi_{q} - \varphi_{b})}\langle \nu_0 |\kappa , k_z , m ,\lambda\rangle \frac{\kappa d\kappa }{(2\pi)^2}\frac{dk_z}{(2\pi)}.
\end{aligned}
\end{equation}
Here $\langle \nu_0|\kappa, k_z, m,\lambda\rangle \propto \delta(k_{\nu_0,\perp} - \kappa)\delta(k_{\nu,z} - k_z)e^{im\varphi_{q}}\delta_{s_{\nu_0}
,\lambda}$ is the wave function of a Bessel photon \cite{Knyazev2018Feb}. A Bessel vortex photon $|\kappa,k_z,m_{\gamma},\lambda\rangle$ possesses a definite value of TAM $m_{\gamma}$ (that consists of OAM and polarization $\lambda$) and definite absolute values of transverse, $\kappa$, and longitudinal, $k_z$, momenta.

A reader may well ask why one would use using a more intricate Bessel-Gaussian wave packet instead of a plain and simple Bessel vortex photon? The limitation of the latter is that it is non-normalizable and, thus, leads to unphysical effects, such as population inversion at $t \gg \Gamma^{-1}$. Indeed, for a Bessel incident photon, Eq. \eqref{Ce} results in the excited state coefficient oscillating at large times with $\lim_{t \rightarrow \infty} \sum\limits_{\nu}|C_{\nu,n}(t)|^2 \neq 0$. Therefore, we describe the incident photon as a localized Bessel-Gaussian wave packet with central transverse $\kappa_{\text{c}}$ and longitudinal $k_{\text{c}}$ momenta. For simplicity we assume equal uncertainties of momenta, \[ \sigma_\kappa = \sigma_{z} = \sigma . \]

For a vortex incident photon approaching the atom at a fixed impact parameter $\bm{b}$, the probability amplitudes can straightforwardly be obtained via Eqs. \eqref{Ce}, \eqref{nu0}:
\begin{equation}
\label{pr_a}
\begin{aligned}
    &C^{\text{v}}_{e,\nu,n}(t) = \sum\limits_{\nu_0}\langle \nu_0 | \gamma_i\rangle C^{\text{pw}}_{e,\nu,n}(t),\\
    &C^{\text{v}}_{g,\nu_1,\nu_2}(t) = \sum\limits_{\nu_0}\langle \nu_0 | \gamma_i\rangle C^{\text{pw}}_{g,\nu_1,\nu_2}(t),
\end{aligned}
\end{equation}
where ''v'' stands for ''vortex''.
In an experiment, however, the position of the atom cannot be controlled with arbitrary precision. Instead, the incident photon interacts with a localized atomic target described by some distribution function $n(\bm{b})$. For further analysis we assume Gaussian distribution of the atom in the trap
\begin{equation}
\label{atomic distribution function}
    n(\bm{b}) = \frac{1}{\pi\sigma_{\text{b}}^2}\exp\left(-\frac{b^2}{\sigma_{\text{b}^2}}\right),
\end{equation}
centered on the the $z$-axis with the width $\sigma_{\text{b}}$ determined by the finite size of the localization region of the trap's holding potential and thermal motion of the atom. Note that velocity distribution is neglected here, assuming that the Doppler frequency shifts for sufficiently cold atoms in the trap are much smaller than the natural linewidth.

Modern atomic and ion traps constrain particles in a 100-nm region or even tighter \cite{Bradac2018Jun, Morrissey2013Aug}. On the contrary, wave packets of visible spectrum photons are of $\sigma^{-1} \sim 100$-$\mu$m sizes at least and, hence, much wider than the traps, $\sigma_{\text{b}} \ll \sigma^{-1}$. This condition allows us to evaluate the coefficients \eqref{pr_a} analytically with high precision (for details see Sec. III of the Supplemental Material \cite{supp}).

For further estimates we consider a paraxial vortex photon, well-localized in momentum space around $\kappa = \kappa_{\text{c}}$, $k_z = k_{\text{c}}$, with optical frequencies $\omega_{\text{c}} = \sqrt{\kappa_{\text{c}}^2 + k_{\text{c}}^2} \sim 1 \text{eV}$ and transverse $\kappa_{\text{c}} \approx (0.1 - 0.01) \omega_{\text{c}}$, and longitudinal $k_{\text{c}} \approx \omega_{\text{c}}$ momenta. The momentum uncertainty is taken to be ten to hundred times smaller than the mean transverse momentum, $\sigma \approx (0.1 - 0.01) \kappa_{\text{c}}$.

\section{Results}
\label{sec3}

\begin{figure*}[t]
    \begin{subfigure}{0.32\linewidth}

    \centering
        \includegraphics[height=5.0cm]{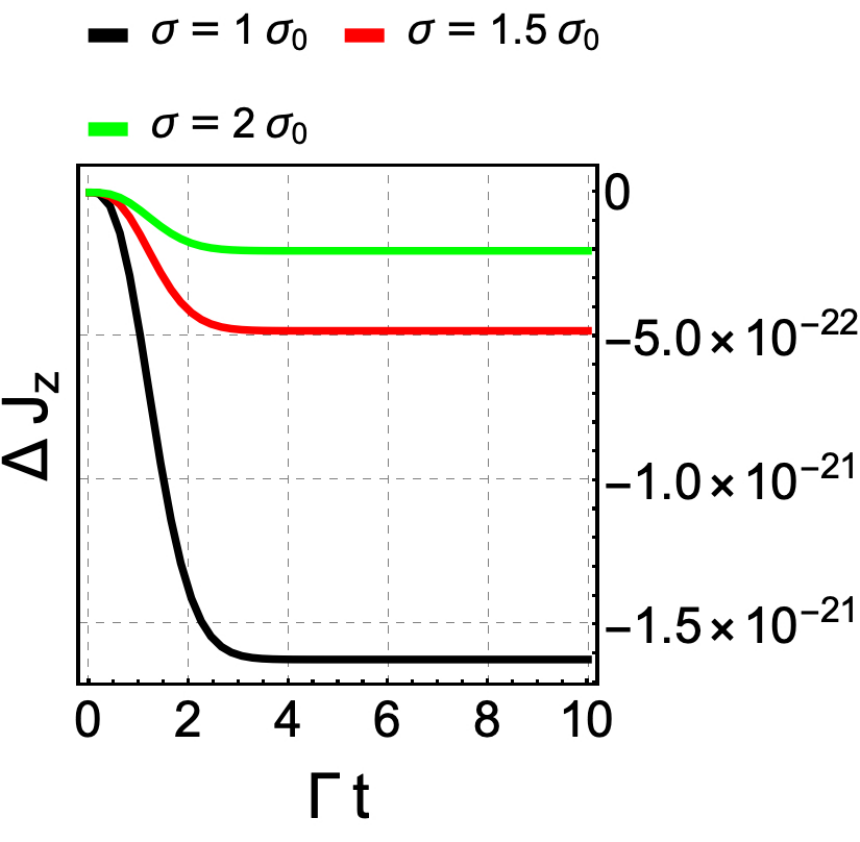}
        \caption{ }
        \label{2a}
    \end{subfigure}
    \hfill
    \begin{subfigure}{0.32\linewidth}

        \centering
        \includegraphics[height=5.0cm]{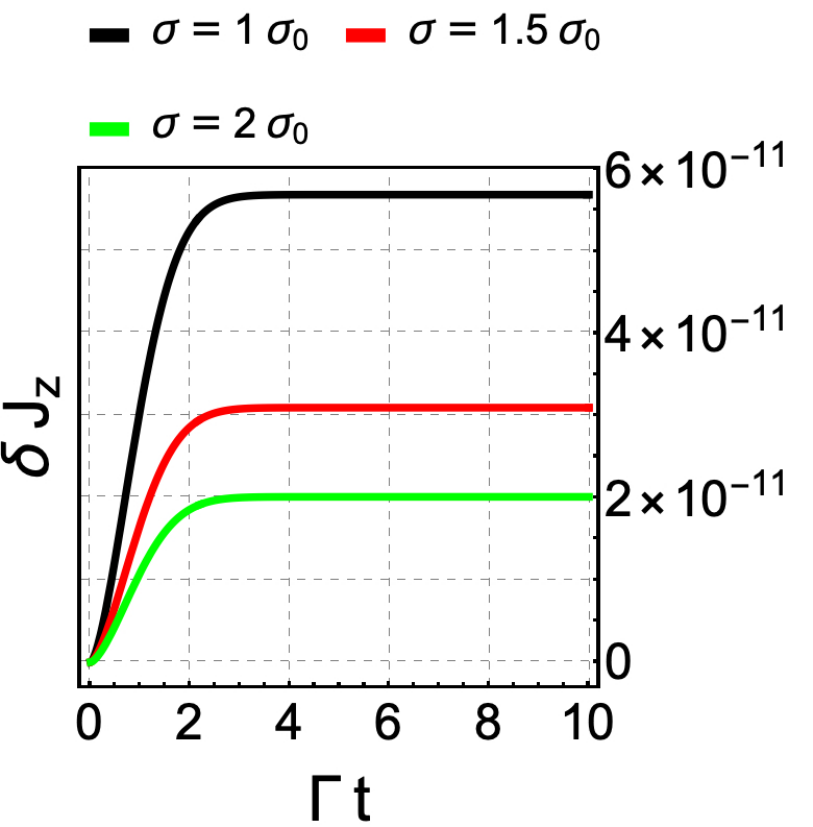}
        \caption{ }
        \label{2b}
    \end{subfigure}
    \hfill
    \begin{subfigure}{0.32\linewidth}

\vspace{-0.1cm}

        \centering
        \includegraphics[height=5.0cm]{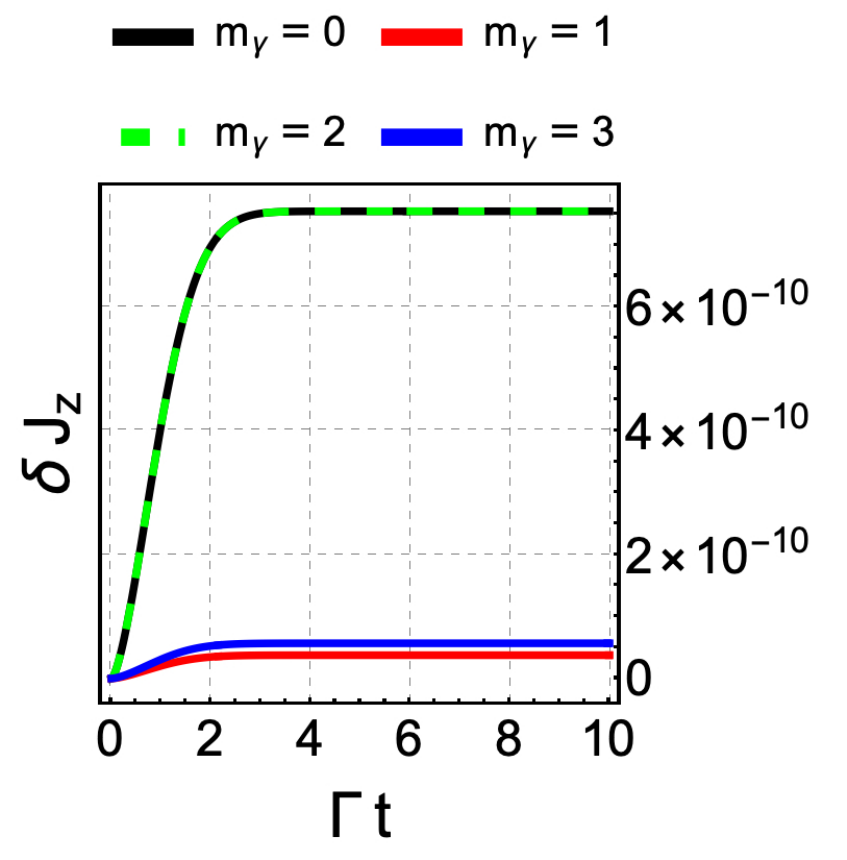}
        \caption{}
        \label{2c}
    \end{subfigure}
    \captionsetup{justification=justified, singlelinecheck=false}
    \caption{\small Time dependence of the TAM mismatch $\Delta J_z$ (a) and the TAM variation $\delta J_z$ (b) for different momentum dispersion of the incident photon: $\sigma = 1\sigma_0$, $1.5\sigma_0$, $2\sigma_0$ where $\sigma_0 = 0.1 \kappa_{\text{c}} = 10^{-2}\omega_{\text{c}} = 2.1 * 10^{-2}$ eV  ( coherence length $\sigma_0^{-1} \approx 9.38 \;\mu m$), $m_{\gamma} = 3$, $m_e = 1$, $\lambda = 1$. (c): Time dependence of the TAM variation for different TAM of the incident photon $m_{\gamma} = 0,1,2,3$ and $\sigma = \sigma_0, m_e = 1, \lambda = 1$. In (a),(b),(c) zero detuning of the incident photon frequency is assumed.}
    \label{Fig:2}
\end{figure*}

The characteristic duration of photon emission and generation of the entangled photon pair is of the order of $\Gamma^{-1}$. Let us now consider the state of the system $|\psi(t) \rangle$, described by Eq. \eqref{st}, with probability amplitudes Eq. \eqref{pr_a}. At sufficiently large times, $t \gg \Gamma^{-1}$, it decouples into the atomic and the electromagnetic field parts:
\begin{equation}
\label{fina two-photon state}
\begin{aligned}
&|\psi(t)\rangle = |\text{g}\rangle|\gamma_f\rangle, \\
&|\gamma_{\text{f}}\rangle = \sum\limits_{\nu_1,\nu_2}C^{\text{v}}_{g,\nu_1,\nu_2}(t)|\nu_1;\nu_2\rangle.
\end{aligned}
\end{equation}

One of our key findings is the state $|\gamma_{\text{f}}\rangle$ being entangled in vortex basis:
\begin{equation}
\label{vortex_basis}
\begin{aligned}
    |\gamma_{\text{f}}\rangle & = \sum\limits_{l_1,l_2 = -\infty}^{\infty}\sum\limits_{\nu_1,\nu_2}C_{\text{g},\nu_1,\nu_2}^{\text{v}}(t)e^{il_1\varphi_{\nu_1} + il_2\varphi_{\nu_2}} \times \\
    & |q_{\nu_1,\perp},q_{\nu_1,z},l_1,s_{\nu_1} ; q_{\nu_2,\perp},q_{\nu_2,z},l_2,s_{\nu_2}\rangle,
\end{aligned}
\end{equation}
where
\begin{equation}
    \begin{aligned}
        &|q_{\nu_1,\perp},q_{\nu_1,z},l_1,s_{\nu_1} ; q_{\nu_2,\perp},q_{\nu_2,z},l_2,s_{\nu_2}\rangle = \\
        & \frac{1}{\sqrt{2}}|q_{\nu_1,\perp},q_{\nu_1,z},l_1,s_{\nu_1} \rangle|q_{\nu_2,\perp},q_{\nu_2,z},l_2,s_{\nu_2}\rangle + \\
        & \frac{1}{\sqrt{2}}|q_{\nu_2,\perp},q_{\nu_2,z},l_2,s_{\nu_2}\rangle|q_{\nu_1,\perp},q_{\nu_1,z},l_1,s_{\nu_1} \rangle.
    \end{aligned}
\end{equation}
Indeed, Eq. (19) from \cite{supp} guarantees $C_{\text{g},\nu_1,\nu_2}^{\text{v}}(t) \ne e^{-il_1'\varphi_{\nu_1}}e^{-il_2'\varphi_{\nu_2}} f(q_{\nu_{1, 2},\perp},q_{\nu_{1, 2},z})$, where $f(q_{\nu_{1, 2},\perp},q_{\nu_{1, 2},z})$ does not depend on azimuthal angles. Hence, the state Eq. \eqref{vortex_basis} cannot be factorized as a tensor product of single photon vortex states.

Note that entangled state \eqref{vortex_basis} will also arise in the case of modes containing more than one photon in the initial state, due to the linear evolution of the system. However, the contributions to the final state arising from these modes can negatively affect the efficiency of generating an entangled pair of twisted photons due to the complication of selecting a suitable pair during detection. In this paper, we do not consider these contributions, assuming that the relative amplitudes of the multiphoton modes in the initial state are small.

Let us consider the probability of detecting a pair of photons with different values of orbital angular momentum. To do this, we average the final state of the photon over the impact parameter \textbf{b} and all quantum numbers except $l_1$ and $l_2$:

\begin{equation}
\label{probability_l1_l2}
\begin{aligned}
    P_{l_1, l_2} &= \int d^2b \ n(\textbf{b}) \times \\ &\sum\limits_{\substack{q_{1,\perp},q_{1,z},s_{1} \\  {q_{2,\perp},q_{2,z},s_{2}}}} | \langle q_{1,\perp},q_{1,z},l_1,s_{1} ; q_{2,\perp},q_{2,z},l_2,s_{2}|\gamma_{\text{f}}\rangle |^2,
\end{aligned}
\end{equation}

\begin{figure}
    \centering    \includegraphics[width=1.\linewidth]{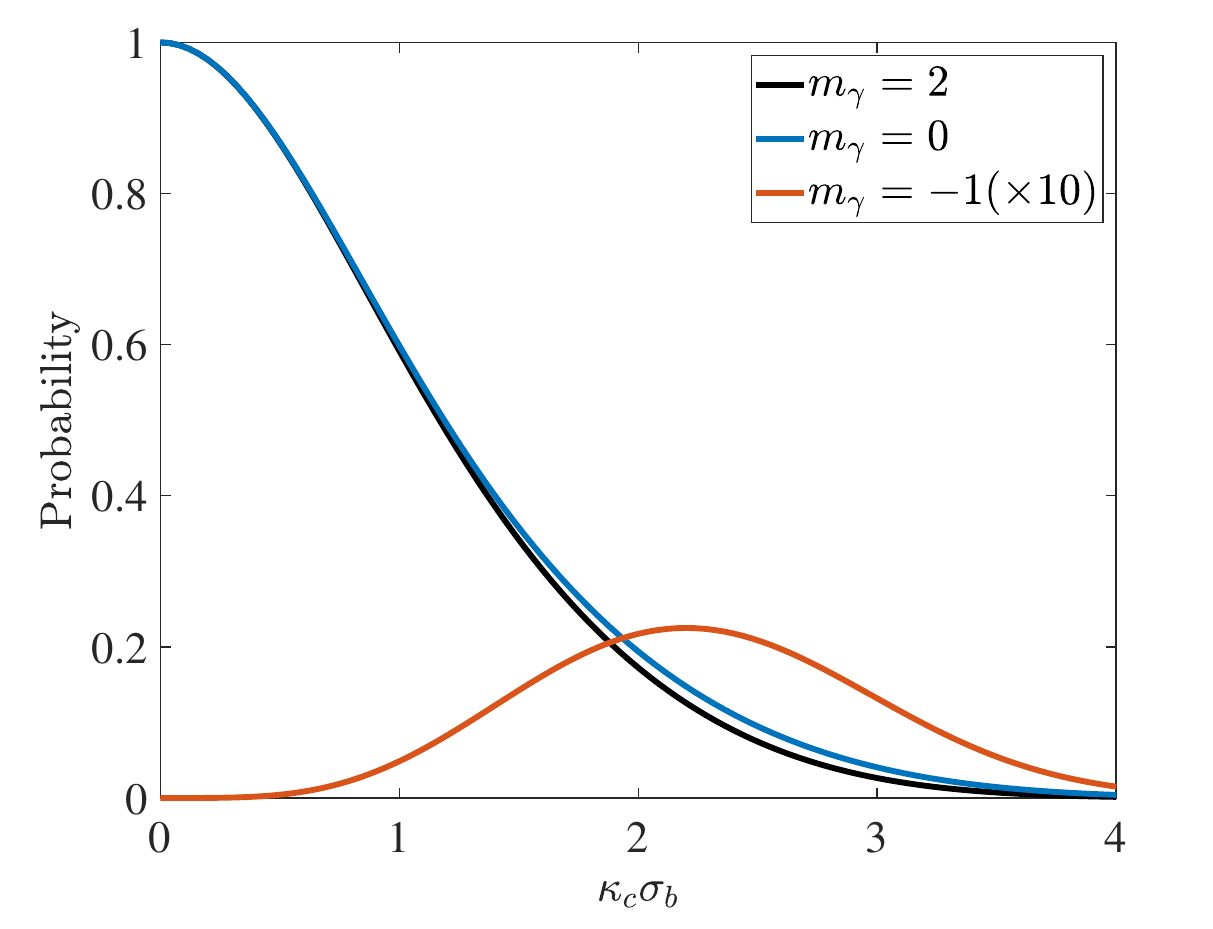}
    \caption{Dependence of probability of detecting an entangled pair with $l_{1(2)}=m_e=1$ and $l_{2(1)}=m_\gamma$ on the spread of the impact parameter $\sigma_b$  for different $m_\gamma$ with $\lambda = 1$, $\sigma=0.1 \kappa_c=10^{-2}\omega_c=10.2 * 10^{-2}$ eV. The graph for $m_\gamma=-1$ is scaled 10 times along the vertical axis for clarity.\textbf{}}
    \label{probability_m0_me}
\end{figure}

As an example, we consider the generation of an entangled photon pair by resonant scattering of an initial photon with $m_\gamma=-1,0,2$ by a hydrogen atom in the 2p state with $m_e=1$. The probability of detecting an entangled pair satisfying the angular momentum conservation law is given by $P_{m_\gamma m_e}+P_{m_e m_\gamma}$, calculated using \eqref{probability_l1_l2} at $t\xrightarrow{}\infty$. Fig. \ref{probability_m0_me} shows the dependence of this probability on the spread of the impact parameter $\sigma_b$. It can be seen that in the case of $m_\gamma=0$ and $2$, the theoretical efficiency of pair generation strongly depends on the size of the localization region of atoms in the trap and varies from zero to unity within the framework of our idealized model. With an increase in the spread of the impact parameter, the angular momentum conservation law is not precisely fulfilled and the contribution of processes associated with the absorption and re-emission of the incident photon through the transition with $m_e=1$ increases. This leads to an increase in the probability of detecting a pair of photons with indiscernible values of moments $l_1=l_2=1$ and a decrease in the efficiency of generating entangled pair with $l_{1(2)}=1$ and $l_{2(1)}=2$. Moreover, in the case of $m_\gamma=-1$, for a certain spread of the impact parameter, the probability of detecting a pair of photons with $l_{1(2)}=1$ and $l_{2(1)}=-1$ has a maximum, since in this range the contribution of the interaction with the transition to a level with a magnetic quantum number of -1 is maximum.

Another factor not included in the present model is the residual population of the ground state in the initial condition. This effect arises mainly from timing jitter between the arrival of the incident photon and the moment of atomic excitation. In the resonant case, such jitter does not alter the normalized characteristics of the final biphoton state, since all temporal components are uniformly rescaled by a common factor that cancels upon normalization. However, it reduces the overall probability of generating an entangled pair due to the finite likelihood of the field remaining in a single-photon state. This reduction is negligible provided that the average jitter is much shorter than the excited-state lifetime.

We note that in the present treatment the atomic center of mass recoil has been neglected, assuming the atom to be tightly confined in a trap. In reality, the transfer of transverse momentum from the incident photon leads to a small recoil of the atom, which carries away part of the total angular momentum. This effect reduces the purity of the biphoton entangled state, especially in the high-$m_\gamma$ regime, where the transverse momentum component $\kappa_c$ becomes larger. For optical photons interacting with heavy atoms or having the small $\kappa_c$, however, the recoil energy $\kappa_c^2/2M$ is many orders of magnitude smaller than the natural linewidth $\Gamma$, and the corresponding entanglement  degradation can safely be neglected. A more detailed analysis of recoil effects in twisted light-induced processes can be found in~\cite{afanasev2021recoil}.

We finally note that the OAM-entangled biphoton state predicted in Eq. (12) can be detected using  established experimental techniques. In particular, orbital angular momentum modes can be resolved  by OAM sorters based on log-polar coordinate transformations, which map different $l$ values into  spatially separated output channels. Alternatively, interferometric schemes employing Dove prisms or  Mach–Zehnder interferometers allow discrimination of superpositions of OAM states, while projective  measurements with spatial light modulators or $q$-plates can be used to analyze individual modes.  Coincidence detection in such measurement bases enables direct verification of the biphoton  entanglement in the OAM degree of freedom, as demonstrated in previous experimental works  on twisted photon entanglement \cite{dada2011experimental,malik2016multi}.

\subsection{Mean TAM and Variance}

The state $|\gamma_{\text{f}}\rangle$ carries all the information about the entangled pair of photons, including the mean value and the variance of TAM:
\begin{equation}
\label{lo}
\begin{aligned}
    &J_{z} = \int d^2b \, n(\bm{b})\frac{\langle  \gamma_{\text{f}} | \hat{J}_z | \gamma_{\text{f}} \rangle}{\langle \gamma_{\text{f}}|\gamma_{\text{f}}\rangle }, \\
    &(\delta J_{z})^2 = \int d^2b \, n(\bm{b})\left[\frac{\langle \gamma_{\text{f}} | \hat{J}_z^2 | \gamma_{\text{f}} \rangle}{\langle \gamma_{\text{f}}|\gamma_{\text{f}}\rangle }-\frac{\langle  \gamma_{\text{f}} |\hat{J}_z | \gamma_{\text{f}} \rangle^2}{ \langle \gamma_{\text{f}}|\gamma_{\text{f}}\rangle^2}\right].
\end{aligned}
\end{equation}

Here $\hat{J}_z = \hat{J}_z^{(1)} + \hat{J}_z^{(2)}$ is the sum of the $z$ - projection TAM operators acting on each photon.  In Eq. \eqref{lo}, we account for the uncertainty in the position of the atom and average the expectation values with the atomic distribution function \eqref{atomic distribution function}. The explicit expressions for the average TAM and the variance are
\begin{equation}
\label{mean value and var}
\begin{aligned}
    &J_z = m_{\gamma} + m_e + \\
    & \sum\limits_{n}\frac{(n-m_{\gamma})}{[(|m_{\gamma}-n|)!]^2}\left(\frac{\kappa_{\text{c}}\sigma_{\text{b}}}{2}\right)^{2|m_{\gamma}-n|}\mathcal{I}_{n,\lambda}(\Gamma,\omega_{\text{c}},\kappa_{\text{c}},\sigma,t),\\
    &(\delta J_z)^2 = \\
    & \sum\limits_{n}\frac{(n-m_{\gamma})^2}{[(|m_{\gamma}-n|)!]^2}\left(\frac{\kappa_{\text{c}}\sigma_{\text{b}}}{2}\right)^{2|m_{\gamma}-n|}\mathcal{I}_{n,\lambda}(\Gamma,\omega_{\text{c}},\kappa_{\text{c}},\sigma,t),
\end{aligned}
\end{equation}
where $\mathcal{I}_{n,\lambda} (\Gamma,\omega_c,\kappa_c,\sigma,t)$ arises as a weight factor originating from the time
integration of the excitation--emission amplitudes (see Supplemental Material, Eq.~(27)). 
Physically, it characterizes the probability of TAM transfer through the channel with magnetic quantum  number $n$ and photon polarization $\lambda$, taking into account the finite lifetime of the excited  state ($\Gamma^{-1}$), the spectral width of the photon wave packet ($\sigma$), and the paraxial geometry of the incident vortex photon. The expression is derived under the Markov approximation and the condition of tight atomic localization ($\sigma_b \ll \sigma^{-1}$), which allow us to expand the Bessel functions and evaluate the integrals analytically. In this sense, $\mathcal{I}_{n,\lambda}$ encapsulates the dynamical and spectral response of the atom--photon system. A detailed derivation of the the average value and the variance for a fixed impact parameter $\bm{b}$ is given in Sec. IV of the Supplemental Material \cite{supp}, and the averaging the atomic distribution function is discussed in Sec. V of the Supplemental Material \cite{supp}.

Notice that Eq. \eqref{mean value and var} indicates a mismatch, $\Delta J_z = J_z - (m_{\gamma} + m_e)$, between the mean TAM of the entangled pair of photons and the combined angular momentum of the incident photon and the excited atom --- $m_{\gamma} + m_e$. This is because Hamiltonian \eqref{int} only accounts for the relative motion of the electron and the discrepancy in the average TAM is attributed to the angular momentum transferred to the center of mass motion \cite{afanasev2021recoil}. Classically, this can be explained by a torque acting on the atom due to a non-zero arm $\bm{b}$. Moreover, a non-zero impact parameter breaks the azimuthal symmetry of the wave function of the system. Therefore, $|\gamma_{\text{f}} \rangle $ is not an eigenstate of $\hat{J}_z$ and possesses a finite variation of TAM.  At the same time, in the limit $\sigma_{\text{b}} \rightarrow 0$ corresponding to a central collision both the mismatch $\Delta J_z$ and variation $\delta J_{z}$ vanish.

For definiteness, we further consider the $3p - 3s$ transition in Na with the resonance frequency $\omega_{a} = 2.1$ eV. The lifetime of the excited state, $16.4$ ns, corresponds to the decay rate (spectral width) $\Gamma = 4* 10^{-8}$ eV; coupling factor can be estimated from the oscillator strength \cite{theodosiou1988accurate}. We assume a zero detuning of the incident photon central frequency from the atomic resonance, $
\Delta_{\text{c}} = \omega_{a} - \omega_{\text{c}} = 0$, and take $\kappa_{\text{c}} = 0.1 \omega_{\text{c}}$. The mean value mismatch and the variation of the angular momentum exhibit minimal dependence on the detuning, only decreasing slightly for off-resonant incident photons. The polarization of the incident photon $\lambda = 1$ is taken, and for the other polarization, the results are qualitatively the same. %\sout{For definiteness,} we consider the atom to be initially in a $p$-orbital excited state with $m_e = 1$ and the ground state of the atom is an $s$-orbital. 
The choice of excited state is not crucial; assuming a higher-OAM excited level only alters the range of values for the magnetic quantum number $n$. To ensure validity of the two-level approximation, we select a transition typically corresponding to the smallest energy gap. Likewise, it is not necessary to assume an $s$-orbital ground state, although such a choice eases the calculations. The atom localization width is $\sigma_{\text{b}} = 100 \; \text{nm}$.

\begin{figure}
    \centering
    \includegraphics[width=1.\linewidth]{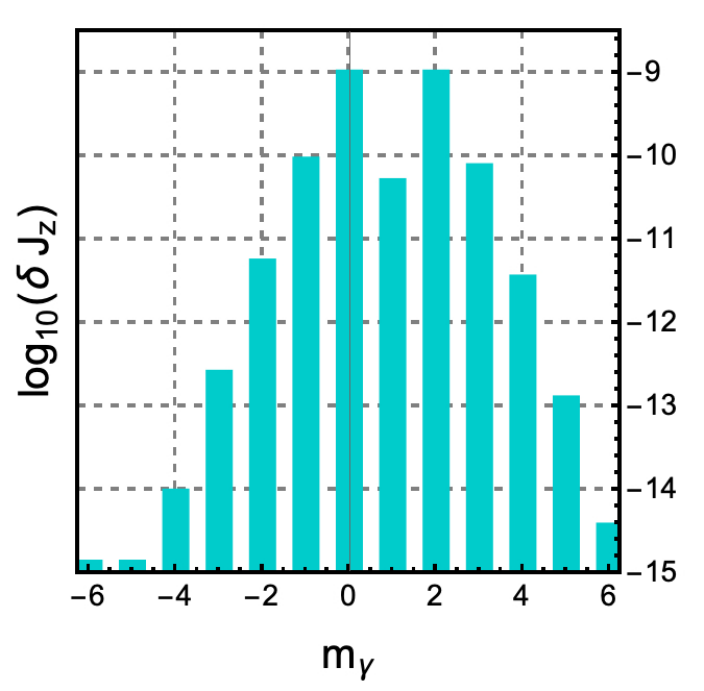}
    \caption{Dependence of the TAM variation on the TAM of the incident photon $m_{\gamma}$ with $\sigma = \sigma_0, m_e = 1, \lambda = 1, t = 10\Gamma^{-1}$.}
    \label{fig:3}
\end{figure}

The time dependence of the $\Delta J_z$ and $\delta J_z$ is shown in in Fig. \ref{Fig:2}. Despite the plots starting at $t = 0$, the dependencies posses the physical meaning of the mismatch and the variation of TAM only for $t \gg \Gamma^{-1}$ when the entangled pair is already created. In this region, $\Delta J_z$ (Fig. \ref{2a}) and $\delta J_z$ (Fig. \ref{2b}) are practically constant. The black, red, and green lines correspond to different values of the incident photon momentum dispersion, $\sigma = \{ 1\sigma_0$, $1.5\sigma_0$, $2\sigma_0 \}$, where $\sigma_0 = 0.1\kappa_{\text{c}}$. The mismatch $\Delta J_z$ increases for a larger momentum dispersion, $\sigma$, although its value is negligibly small compared to the combined angular momentum of the initial system, $|\Delta J_{z}| \ll m_{\gamma} + m_{e}$. The variation $\delta J_z$, on the other hand, decreases for a larger momentum dispersion and its characteristic value is of the order of $10^{-7} - 10^{-8}$ which is also small, $\delta J_z \ll m_{\gamma} + m_{e}$.

In Fig. \ref{2c} we compare the TAM variation of the entangled photon pair for different TAM of the incident photon. The variation exhibits two maxima, for $m_{\gamma} = 0$ and $m_{\gamma} = 2$, which can be seen directly from Eq. \eqref{mean value and var}. For the parameters used the quantity $\kappa_{\text{c}}\sigma_{\text{b}}/2 = 0.053 \ll 1$, and therefore the variation is maximum whenever $|m_{\gamma} - n| = 1$. As we assume $p$-orbital excited state, the magnetic index $n$ runs over the values $\{-1,0,1\}$ and, hence, the condition above can only be met if $|m_{\gamma}| \le 2$. Depending on the TAM of the incident photon the greatest contribution to the variation comes from different $n$:
\begin{table}[h]
\begin{adjustbox}{width=0.3\columnwidth}
\begin{tabular}{|c|cc|cc|cc|}
\hline
$n$          & \multicolumn{2}{c|}{$-1$}   & \multicolumn{2}{c|}{$0$}    & \multicolumn{2}{c|}{$1$}   \\ \hline
$m_{\gamma}$ & \multicolumn{1}{c|}{-2} & 0 & \multicolumn{1}{c|}{-1} & 1 & \multicolumn{1}{c|}{0} & 2 \\ \hline
\end{tabular}
\end{adjustbox}
\end{table}
\newline
In addition, we note $\mathcal{I}_{n,\lambda}(\Gamma,\omega_{\text{c}},\kappa_{\text{c}}, \sigma,t) \sim (d_{n,\lambda}^{1}(\theta_{\text{c}}))^2$ (see \cite{Scholz-Marggraf2014Jul} and Sec. IV of the Supplemental Material \cite{supp}) , where $\theta_{\text{c}} = \arctan(\kappa_{\text{c}}/\omega_{\text{c}})$ is the central conical angle of the incident photon. A small value of $\theta_{\text{c}} \approx 0.09$ rad leads to $d_{n \ne \lambda,\lambda}^{1}(\theta_{\text{c}}) \ll d_{\lambda,\lambda}^{1}(\theta_{\text{c}})$. As we consider $\lambda = 1$, we find that only $m_{\gamma} = 0 (n = 1)$ and $m_{\gamma} = 2 (n = 1)$ provide the greatest  contribution to the TAM variation in accordance with (Fig. \ref{2c}). Alternatively, for the incident photon with $\lambda = -1$, one would observe maximum TAM variation for $m_{\gamma} = -2$ and $m_{\gamma} = 0$.

In Fig. \ref{fig:3} we demonstrate the TAM variation rapidly dropping for larger $m_{\gamma}$. 
Indeed, for $m_{\gamma} \gg 1$ the magnetic index in the exponent and prefactor in Eq. \eqref{mean value and var} can be neglected and we find
\begin{equation}
\label{ed}
\begin{aligned}
    \delta J_z \approx \left(\frac{\kappa_{\text{c}}\sigma_{\text{b}}}{2}\right)^{m_{\gamma}}\sqrt{\left(\frac{m^2_{\gamma}}{m_{\gamma}!}\right)\sum\limits_{n}\mathcal{I}_{n,\lambda}(\Gamma,\omega_{\text{c}},\kappa_{\text{c}},\sigma,t)}.
\end{aligned}
\end{equation}
In agreement with Fig. \ref{2c} $\ln(\delta J_z)$ exhibits maxima at $m_{\gamma} = \{0, 2\}$ and then decreases linearly with the increase in the incident photon TAM.

\section{Conclusion and outlook}

In this work, we have developed a theoretical framework for describing atomic emission induced by an arbitrary photon wave packet. We have applied this formalism to investigate the interaction between an incident vortex photon and a localized atomic target, with the main focus on the entangled pair of photons generated through this process. Particularly, we have analyzed the transfer of the angular momentum (AM) from the incident photon to the entangled pair. It has been shown that for an experimentally feasible scenario the AM variation of the entangled pair is much smaller than $\hbar$ and sharply drops with the increase in AM of the incident photon. Our findings indicate that the induced emission process allows one to generate entangled pairs of photons with a definite value of AM.

We have proposed a way to generate entangled pairs of vortex photons as an alternative to the commonly used methods, such as spontaneous parametric down conversion. The technique based on the induced emission has the following advantage. The influence of the uncertainty in the position of the atom in a real experiment on the AM transfer can be mitigated by parameters of the incident light. For an atomic localization smaller than incident photon transverse coherence length, $\kappa_{\text{c}}^{-1}$, one can always find a value of AM of the incident photon such that the AM variation of the entangled pair is negligibly small. For atomic localization on the scale of hundreds of nanometers, an incident photon with $m_{\gamma} = 3\hbar$ and the transverse momentum $\kappa_{\text{c}} = 0.1 \;\text{eV}$ leads to a $10^{-11}\hbar$ variation in AM of the pair, whereas for a $10 \mu\text{m}$-localization, $m_{\gamma} = 6  \hbar$ provides roughly the same variation.

In this paper we have focused on the final biphoton state at infinite times. However, the approach we have applied allows one to analyze the time dynamics of the system and the observables. On top of that our results can be readily generalized to any arbitrary incident photon wave packet. Additionally, we note that, in principle, this scheme is applicable for generating entangled pairs of vortex $\gamma$-photons emitted by an excited nucleus induced by an incident vortex $\gamma$-photon.

We suggest three possible extensions of the current study. First, one can use more realistic atomic models including fine structure of atomic levels. Second, an analysis beyond the two-level approximation may enrich the diversity of generated photonic states. Third, from an experimental perspective, it is instructive to account for a delay between the pump and the incident photon.

The possibility to generate entangled vortex photons is of great importance for quantum information and quantum cryptography. The orbital degree of freedom of light can be utilized to enhance the information transfer capacity compared to the traditional protocols relying on the polarization states of light. Quantum entanglement in its turn is key to ensure the security of information transfer. Moreover, it has been shown that the quantum key distribution schemes based on angular momentum of light are more resilient against eavesdropping attacks \cite{article}.

\section*{Funding}

The derivation of the population coefficients in the emission induced by a plane wave in Sec.~\ref{subsec21} are supported by the Foundation for the Advancement of Theoretical Physics and Mathematics “BASIS”. The extension of these calculations to an incident vortex photon in Sec.~\ref{subsec22} is supported by the Ministry of Science and Higher Education of the Russian Federation (Project FSER-2025-0012). The studies of time dynamics of OAM of the entangled pair in Sec. \ref{sec3} are supported by the Russian Science Foundation (Project No.\,23-62-10026;  https://rscf.ru/en/project/23-62-10026/). The studies of influence of the finite localization region of an atom on its interaction with an incident vortex photons are supported by the Russian Science Foundation (Project No. 25-71-00060).

\section*{Acknowledgements}

We are grateful to V. Zalipaev, T. Zalialiutdinov, I. Pavlov and A. Chaikovskaia for many fruitful discussions and advice.

\begin{comment}

\section*{Disclosures}

The authors declare no conflicts of interest.

\section*{Data availability}

No data were generated or analyzed in the research.

\section*{Supplemental document}

The Supplemental Material can be found in ref. \cite{supp}.
    
\end{comment}

\bibliography{references}

\end{document}

\preprint{APS/123-QED}

\title{Supplemental Material to "Generating entangled pairs of vortex photons via induced emission"}% Force line breaks with \\

\author{D.\,V.~Grosman}
\email{dmitriy.grosman@metalab.ifmo.ru}
\affiliation{School of Physics and Engineering,
ITMO University, 197101 St. Petersburg, Russia}

\author{G.\,K.~Sizykh}
\email{georgii.sizykh@metalab.ifmo.ru}
\affiliation{School of Physics and Engineering,
ITMO University, 197101 St. Petersburg, Russia}
\affiliation{Petersburg Nuclear Physics Institute of NRC “Kurchatov Institute”, 188301 Gatchina, Russia}

\author{E.\,O.~Lazarev}
\email{egor.lazarev@metalab.ifmo.ru}
\affiliation{School of Physics and Engineering,
ITMO University, 197101 St. Petersburg, Russia}

\author{G.\,V.~Voloshin}
\email{gavriilvsh@gmail.com}
\affiliation{Peter the Great St. Petersburg Polytechnic University, 195251 St. Petersburg, Russia}

\author{D.\,V.~Karlovets}
\email{dmitry.karlovets@metalab.ifmo.ru}
\affiliation{School of Physics and Engineering,
ITMO University, 197101 St. Petersburg, Russia}
\affiliation{Petersburg Nuclear Physics Institute of NRC “Kurchatov Institute”, 188301 Gatchina, Russia}

\maketitle

\section{Equation for the ground state probability amplitude}
\label{App_GroundState}

Let us consider a Schr\"{o}dinger equation
\begin{equation}
\label{int}
    i\partial_{t} | \psi(t) \rangle =  \sum\limits_{\nu,n}\left[ g_{\nu,n}^* \hat{\sigma}_{+,n}\hat{a}_{\nu}e^{i\Delta_{\nu}t} + \text{h.c.} \right]| \psi(t) \rangle.
\end{equation}
The ground state $|\text{g} \rangle$ of the atom is assumed to be an $s-$orbital and for the excited state we take into account the magnetic sublevels $|\text{e}_{n}\rangle$, labeled by $n$. In Eq. \eqref{int} we imply the rotating wave approximation and $\nu = \{\bm{k}_{\nu},\lambda_{\nu}\}$ is the multi-index containing photon momentum $\bm{k}_{\nu}$ and polarization $\lambda_{\nu}$, $\hat{\sigma}_{+,n} = |\text{e}_n\rangle\langle \text{g}|$, $\hat{a}_{\nu}$ is the photon annihilation operator, $\Delta_{\nu} = \omega_{\nu} - \omega_{\text{a}}$ is the frequency detuning of the mode $\nu$ from the atomic resonance $\omega_{\text{a}}$. To properly account for the interaction between the atom and a vortex photon we consider the exact non-dipolar coupling:
\begin{equation}
\label{coup}
    g_{\nu,n} = -\frac{e}{m}\sqrt{\frac{2\pi}{V\omega_{\nu}}}\langle \text{g}| \bm{e}_{\bm{k},s}\cdot\hat{\bm{p}}e^{i\bm{k}\cdot\bm{r}}| \text{e}_{n} \rangle,
\end{equation}
where $V$ is the normalization volume.

The interaction Hamiltonian in the r.h.s. of this equation changes the number of quanta of light by one, hence, the natural anzats for the solution of the Schr\"{o}dinger equation is
\begin{equation}
\label{st}
    |\psi(t) \rangle = \sum\limits_{\nu,n} C_{e,\nu,n}(t)|\text{e}_{n}\rangle|\nu\rangle + \sum\limits_{\nu_1,\nu_2}C_{g,\nu_1,\nu_2}(t)|\text{g}\rangle|\nu_1;\nu_2\rangle.
\end{equation}
Substituting this into the Schr\"{o}dinger equation results in the following system of equations:
\begin{equation}
\label{Sys}
    \begin{aligned}
        &i\dot{C}_{e,\nu,n}(t) = \sum\limits_{\nu'}\left[ C_{g,\nu',\nu}(t) + C_{g,\nu,\nu'}(t) \right] g_{\nu',n}^* e^{-i \Delta_{\nu'} t}, \\
        &i\dot{C}_{g,\nu_1,\nu_2}(t) = \frac{1}{2}\sum\limits_{n}C_{e,\nu_1,n}(t)g_{\nu_2,n}e^{-i\Delta_{\nu_2}t} + \nu_1 \Leftrightarrow \nu_2.
    \end{aligned}
\end{equation}
Then, one integrates the second equation with the initial condition $C_{g,\nu_1,\nu_2}(0) = 0$ and substitutes the result into the first equation to get the integro-differential equation for the excited state probability amplitude:
\begin{equation}
\label{diffuclt equation}
    \dot{C}_{e,\nu,n}(t) = - \sum\limits_{\nu',n'}\int\limits_{0}^{t}d\tau\big[C_{e,\nu,n'}(\tau)g_{\nu',n'}e^{-i\Delta_{\nu'}t} + C_{e,\nu',n'}(t)g_{\nu,n'}e^{-i\Delta_{\nu}t} \big] g_{\nu',n}^* e^{i\Delta_{\nu'}t}.
\end{equation}
After solving this equation and finding the excited-state coefficient, the ground-state coefficient can be found as
\begin{equation}
\label{eqfg}
    C_{g,\nu_1,\nu_2}(t) = -\frac{i}{2}\sum\limits_{n}\int\limits_{0}^{t}C_{e,\nu_1,n}(\tau)g_{\nu_2,n}e^{-i\Delta_{\nu_2}t\tau}d
    \tau+ \nu_1 \Leftrightarrow \nu_2.
\end{equation}

\section{Excited state probability amplitude for an incident plane wave}
\label{App_CeDerivation}

To solve Eq. \eqref{diffuclt equation}, let us introduce an auxiliary function
\begin{equation}
    \mathcal{C}_{n,n'}(t,t') = \sum\limits_{\nu}g_{\nu,n}^{*} e^{i\Delta_{\nu}t'}C^{\text{pw}}_{e,\nu,n'}(t).
\end{equation}
It allows us to rewrite Eq. \eqref{diffuclt equation} as
\begin{equation}
\label{auxeqeq}
    \dot{C}^{\text{pw}}_{e,\nu,n}(t) = -\frac{\Gamma}{2}C^{\text{pw}}_{e,\nu,n}(t) - \int\limits_{0}^{t}\sum\limits_{n'}g_{\nu,n'}\mathcal{C}_{n,n'}(\tau,t)e^{-i\Delta_{\nu} \tau} d\tau.
\end{equation}
Here we have used (see \cite{Scholz-Marggraf2014Jul})
\begin{equation}
\label{gandG}
g_{\nu,n} = \exp{(in\varphi_{\nu})} G_n(\omega_{\nu},\theta_{\nu}),
\end{equation}
as well as the Markov approximation for the first term in Eq. \eqref{diffuclt equation}. Here $\varphi_{\nu}$ is the azimuthal angle of the photon wave vector corresponding to the mode $\nu$, \begin{equation}
\begin{aligned}
    G_n(\omega_{\nu},\theta_{\nu})= -\frac{e}{m}\sqrt{\frac{2\pi}{V\omega_{\nu}}}\sum\limits_{n'}d^{1}
    _{n,n'}(\theta_\nu)M_{n',0}^{0}, 
\end{aligned}
\end{equation}
where the quantity $d^1_{n,n'}(\theta)$ is the Wigner's (small) $d$-matrix element that describes how the components of the angular momentum state of the system change when the coordinate system rotates the z-axis by an angle $\theta$ \cite{varshalovich1988quantum}.
The matrix element $M_{n,0}^{0}$ for the excitation of a hydrogenlike atom from its $s$-state can be written as

\begin{equation}
\begin{aligned}
M_{n,0}^{0}=&-i\frac{e}{m} \sum_{L} i^L \sqrt{\frac{(2 L+1)^3}{3}} \\
& \times \langle 1,0,L,0|1 0\rangle \langle 1,\lambda,L,0|1 n \rangle \\
& \times \int_0^{\infty} R_{n_e,1}(r) j_L(kr) \frac{\partial{R_{n_g,0}(r)}}{\partial{r}}r^2dr,
\end{aligned}
\end{equation}
where $R_{n_{e(g)},1}$ is the radial wave function of the hydrogenlike atom whith principal quantum number $n_{e(g)}$ and azimuthal quantum number equal to $1$.

The Markov approximation can mathematically be stated as
\begin{equation}
\label{mark}
\sum\limits_{\nu}|g_{\nu,n}|^2e^{i\Delta_{\nu}(t - \tau)} = \Gamma \delta(t - \tau),
\end{equation}
where $\Gamma$ is the decay rate of the excited state.

Next, multiplying both sides of Eq. \eqref{auxeqeq} by $g_{\nu, r}\exp{[i(\omega_{\text{a}} - \omega_{\nu})t']}$ and summing over $\nu$ one obtains
\begin{equation}
\label{auxeqeq1}
    \partial_t \mathcal{C}_{r,n}(t,t') = -\frac{\Gamma}{2}\mathcal{C}_{r,n}(t,t') -\int\limits_{0}^{t}\sum\limits_{n',\nu}|G_{n'}(\omega_{\nu},\theta_{\nu})|^2e^{i(n' -r)\varphi_{\nu}} \mathcal{C}_{n,n'}(\tau,t)e^{i\Delta_{\nu}(t'-\tau)}d\tau.
\end{equation}
We also apply the Markov approximation to the integral in the r.h.s. of this equation. Notice that Eq. \eqref{auxeqeq} implies that only $\mathcal{C}_{n,n'}(\tau,t)$ with arguments $t > \tau$ contributes to $C_{e,\nu,n}(t)$. Therefore, the second term in Eq. \eqref{auxeqeq1} vanishes for $t' > t$. This is because of the delta function $\delta(t' - \tau)$ that arises from the Markov approximation \eqref{mark}. Indeed, for $t' > t$, the argument of the delta function is always non-zero in the integration region. Thus, Eq. \eqref{auxeqeq1} turns into
\begin{equation}
\label{s}
    \partial_t \mathcal{C}_{r,n}(t,t') =
    -\frac{\Gamma}{2}\mathcal{C}_{r,n}(t,t').
\end{equation}

The initial condition for the excited state probability amplitude, $C^{\text{pw}}_{e,\nu,n}(0) = \delta_{n,m_e}\delta_{\nu,\nu_0}$, implies the initial condition for the auxiliary function: $\mathcal{C}_{n,n'}(0,t') = g_{\nu_0,m_e}^*e^{i\Delta_{\nu_0}t'}$. The solution to Eq. \eqref{s} satisfying this initial condition reads
\begin{equation}
\label{Ca}
    \mathcal{C}_{n,n'}(t,t') = g_{\nu_0,n}^* \delta_{n',m_e}e^{-\frac{\Gamma}{2}t + i\Delta_{\nu_0}t}.
\end{equation}

Finally, one substitutes Eq. \eqref{Ca} into Eq. \eqref{auxeqeq} to obtain an ordinary differential equation for the excited state probability amplitude. Integrating it over time leads to
\begin{equation}
\label{Ce}
\begin{aligned}
    C^{\text{pw}}_{e,\nu,n}(t) = &\delta_{\nu,\nu_0}\delta_{n,m_e}e^{-\frac{\Gamma}{2}t} - g_{\nu,m_e}g^{*}_{\nu_0,n} e^{-\frac{\Gamma}{2}t} \int\limits_{0}^{t}dt_2 \int\limits_{0}^{t_2}e^{\frac{\Gamma}{2}t_2 + i\Delta_{\nu_0}t_2 - \frac{\Gamma}{2}t_1 - i\Delta_{\nu}t_1} d t_1.
\end{aligned}
\end{equation}

\section{Excited and ground state probability amplitudes for an incident vortex photon}
\label{App_C}

The state of the system resulting from the interaction of an exited atom with a vortex incident photon is described by
\begin{equation}
    |\psi(t)\rangle = \sum\limits_{\nu,n} C_{e,\nu,n}^{\text{v}}(t)|\text{e}_n\rangle|\nu\rangle + \sum\limits_{\nu_1,\nu_2}C_{g,\nu_1,\nu_2}^{\text{v}}(t)|\text{g}\rangle|\nu_1;\nu_2\rangle.
\end{equation}
Here, $C_{e,\nu,n}^{\text{v}}(t)$ is obtained by integrating the plane wave probability amplitude \eqref{Ce} with the vortex photon wave function, 
\begin{equation}
\label{nu0}
    \langle \nu_0 |\gamma_i \rangle = N \int 
    e^{-(\kappa - \kappa_\text{c})^2/2\sigma_{\kappa}^2}
    e^{ - (k_z - k_{\text{c}})^2/2\sigma_z^2} e^{i\kappa b \cos(\varphi_{q} - \varphi_{b})}\langle \nu_0 |\kappa , k_z , m ,\lambda\rangle \frac{\kappa d\kappa }{(2\pi)^2}\frac{dk_z}{(2\pi)},
\end{equation}
as follows:
\begin{equation}
    C_{e,\nu,n}^{\text{v}}(t) = \sum\limits_{\nu_0}\langle \nu_0 | \gamma_i \rangle C_{e,\nu,n}^{\text{pw}}(t).
\end{equation}

Provided the impact parameter, at which the photon approaches the atom, is small compared to the photon coherence length, $b \ll \sigma^{-1}$, the integration can be performed approximately using Laplace method \cite{murray2012asymptotic}:
\begin{equation}
\label{C_e_e}
\begin{aligned}
&C_{e,\nu,n}^{\text{v}}(t) =\delta_{n,m_e}\langle \nu | \gamma_i \rangle e^{-\frac{\Gamma}{2}t} \\
- N Vi^{m-n}\sigma^2\kappa_{\text{c}}G_n(\omega_{\text{c}},\theta_{\text{c}})e^{i(m-n)\varphi_b} & J_{m-n}(\kappa_{\text{c}} b) g_{\nu,m_e}e^{-\frac{\Gamma}{2}t}\int\limits_{0}^{t}\int\limits_{0}^{t_1}e^{\frac{\Gamma}{2}t_1 - \frac{\Gamma}{2}t_2 - i\Delta_{\nu} t_2 + i\Delta_{\text{c}} t_1 -\frac{\sigma^2t_1^2}{2}}dt_1dt_2.
\end{aligned}
\end{equation}
Here, $\Delta_c=\omega_\text{a} - \omega_\text{c}$, $\omega_c = \sqrt{\kappa_\text{c}^2 + k_{\text{c}}^2}$, $\theta_{\text{c}} = \tan^{-1}(\kappa_{\text{c}}/k_{l,c})$, and $N$ is the normalization constant.

The ground state probability amplitude is connected with the excited state one via expression \eqref{eqfg} and is given by
\begin{equation}
\label{C_g_g}
    C^{\text{v}}_{g,\nu_1,\nu_2}(t) = g_{\nu_1,m_e}f_n(\omega_{\nu_1},t)\langle \nu_2 | \gamma_i\rangle
- \sum\limits_{n}g_{\nu_1,n}g_{\nu_2,m_e}F_{m,n}(\omega_{\nu_1},\omega_{\nu_2},t) + \nu_1 \Leftrightarrow \nu_2,
\end{equation}
where
\begin{equation}
\label{Ctw}
\begin{aligned}
&f(\omega_{\nu_1},t) = \frac{1 - e^{-\frac{\Gamma}{2}t - i\Delta_{\nu_1}t}}{\Gamma/2 + i\Delta_{\nu_1}},
\end{aligned}
\end{equation}
and
\begin{equation}
\begin{aligned}
    &F_{m,n}(\omega_{\nu_1},\omega_{\nu_2},t) = N Vi^{m-n}\sigma^2\kappa_{\text{c}}G_n(\omega_{\text{c}},\theta_{\text{c}})e^{i(m-n)\varphi_b} \\
    & \times J_{m-n}(\kappa_{\text{c}} b)
\int\limits_{0}^{t}dt_3\int\limits_{0}^{t_3}dt_2\int\limits_{0}^{t_2}dt_1e^{-
    \frac{\Gamma}{2}(t_3+t_1 - t_2)} e^{-i\Delta_{\nu_1} t_3 + i\Delta_{\text{c}} t_2- i\Delta_{\nu_2}t_1 - \frac{\sigma^2t_2^2}{2}}.
\end{aligned}
\end{equation}

\section{The mean value and the variation of TAM}
\label{D}

The matrix elements of the operators $
\hat{J}_z$ and $\hat{J}_z^2$ evaluated in the plane wave basis read
\begin{equation}
\label{ev}
\begin{aligned}
    \langle \nu_1 | &\hat{J}_z | \nu_2 \rangle = \frac{(2\pi)^5}{k_{\nu_1,\perp}^2}(-i)\delta(k_{\nu_1,\perp} - k_{\nu_2,\perp}) \delta(k_{\nu_1,z} -k_{\nu_2,z})\delta_{s_{\nu_1},s_{\nu_2}}\delta'(\varphi_{\nu_1} - \varphi_{\nu_2}),\\
    \langle \nu_1 | &\hat{J}_z^2 | \nu_2 \rangle = -\frac{(2\pi)^5}{k_{\nu_1,\perp}^2}\delta(k_{\nu_1,\perp} - k_{\nu_1,\perp}) \delta(k_{\nu_1,z} -k_{\nu_2,z})\delta_{s_{\nu_1},s_{\nu_2}}\delta''(\varphi_{\nu_1} - \varphi_{\nu_2}),
\end{aligned}
\end{equation}
where $k_{\nu,\perp}, k_{\nu,z}, \varphi_{\nu}$ are absolute value of the transverse, the longitudinal momenta and the azimuthal angle of the photon in the mode $\nu$; $\delta'$ and $\delta''$ denote the first and second derivative of the delta function.

Eqs. \eqref{ev} allow us to evaluate the expectation values of the operators $\hat{J}_z$ and $\hat{J}_z^2$ for the final two-photon state
\begin{equation}
\label{fina two-photon state}
|\gamma_{\text{f}}\rangle = \sum\limits_{\nu_1,\nu_2}C^{\text{v}}_{g,\nu_1,\nu_2}(t)|\nu_1;\nu_2\rangle.
\end{equation}
The expressions are rather cumbersome and we present them in the following manner:
\begin{equation}
\label{exp}
\begin{aligned}
    &\langle \gamma_f |\hat{J}_z | \gamma_f \rangle = (m_{\gamma}+m_e)\langle\gamma_f|\gamma_f \rangle + \sum\limits_{n}(n-m_{\gamma})J_{m_{\gamma}-n}^2(\kappa_{\text{c}} b)\mathcal{I}_{n,\lambda}(\Gamma,\omega_{\text{c}},\kappa_{\text{c}},\sigma,t) , \\
    &\langle \gamma_f |\hat{J}^2_z | \gamma_f \rangle = (m_{\gamma} + m_e) \langle \gamma_f |\hat{J}_z | \gamma_f \rangle + \sum\limits_{n}(n-m_{\gamma})(n+m_0)J^2_{m_{\gamma}-n}(\kappa_{\text{c}} b)\mathcal{I}_{n,\lambda}(\Gamma,\omega_{\text{c}},\kappa_{\text{c}},\sigma,t),
\end{aligned}
\end{equation}
where $\mathcal{I}_{n,\lambda}(\Gamma,\omega_{\text{c}},\kappa_{\text{c}},\sigma,t)$ is given by the six-fold integral
\begin{equation}
\begin{aligned}
    \mathcal{I}_{n,\lambda}(\Gamma,\omega_{\text{c}},\kappa_{\text{c}},\sigma,t) = & [1 + \delta_{n,m_e}] 4N^2V^2\sigma^4\kappa_{\text{c}}^2|G_{n}(\omega_c,\theta_c)|^2\Gamma^2 \\
    \times \int\limits_{0}^{t}dt_3\int\limits_{0}^{t_3}dt_2\int\limits_{0}^{t_2}dt_1\int\limits_{0}^{t}dt'_3 & \int\limits_{0}^{t'_3}dt'_2\int\limits_{0}^{t'_2}dt_1' e^{-
    \frac{\Gamma}{2}(t_3 +t_1 -t_2)  - \frac{\sigma^2t_2^2}{2}}e^{-
    \frac{\Gamma}{2}(t'_3 +t_1' - t_2')  - \frac{\sigma^2t_2'^2}{2}} e^{i\Delta_{\text{c}}(t_2 - t_2')} \\
    & \times \delta(t_3 - t_3')\delta(t_1 - t_1'),
\end{aligned}
\end{equation}
where $G_{n}(\omega_c,\theta_c)$ is connected to $g_{\nu,n}$ via Eq. \eqref{gandG}. The function $\mathcal{I}_{n,\lambda}(\Gamma,\omega_c,\kappa_c,\sigma,t)$ arises as a weight factor in the expectation values of the TAM.  It describes the probability for TAM transfer through the channel with  magnetic quantum number $n$ and photon polarization $\lambda$, taking into account  the finite lifetime of the excited state ($\Gamma^{-1}$), the spectral width of the  incident photon wave packet ($\sigma$), and the paraxial geometry of the vortex photon  ($\kappa_c,\omega_c$).

\section{Averaging of the mean value and the variation of TAM over the impact parameter}
\label{E}

To describe the scenario in which the incident photon interacts with a localized atomic target, one should integrate expressions \eqref{ev} with the transverse atomic distribution function
\begin{equation}
\label{atomic distribution function}
    n(\bm{b}) = \frac{1}{\pi\sigma_{\text{b}}^2}\exp\left(-\frac{b^2}{\sigma_{\text{b}}^2}\right).
\end{equation}
It represents an atom in a trap centered on the the $z$-axis with the width $\sigma_{\text{b}}$. Note, that the dependence of the expectation values on the impact parameter is only due to the Bessel function $J_{m_{\gamma} - n} ( \kappa_{\text{c}} b)$. Recall, that we have previously assumed $b \ll \sigma^{-1}$, which is satisfied if $\sigma_{\text{b}} \ll \sigma^{-1}$. The latter assumption is justified by the fact that for the incident photon of the optical energy range, $\sigma^{-1}$ is of the order $20 \mu\text{m} - 2\text{mm}$, whereas the width of the atomic target can be controlled with tens to hundreds of nanometers precision. Therefore, the Bessel function in Eq. \eqref{ev} can be expanded in power series: $J_{m_{\gamma} - n}(\kappa_{\text{c}} b) \approx (|m_{\gamma} - n|!)^{-1}(\kappa_{\text{c}} b/ 2)^{|m_{\gamma} - n|}$. Integration over the impact parameter with the atomic distribution function then simply yields
\begin{equation}
\label{avaraged with the atomic distribution function}
    \int n(\bm{b}) J^2_{m_{\gamma} - n} (\kappa_{\text{c} }b) d^2b \approx \frac{1}{|m_{\gamma} - n|!}\left(\frac{\kappa_{\text{c}} \sigma_{\text{b}}}{ 2}\right)^{2|m_{\gamma} - n|}.
\end{equation}
With this approximation, Eqs. \eqref{ev} lead to expressions
\begin{equation}
\label{mean value and var}
\begin{aligned}
    &J_z = m_{\gamma} + m_e + \sum\limits_{n}\frac{(n-m_{\gamma})}{[(|m_{\gamma}-n|)!]^2}\left(\frac{\kappa_{\text{c}}\sigma_{\text{b}}}{2}\right)^{2|m_{\gamma}-n|}\mathcal{I}_{n,\lambda}(\Gamma,\omega_{\text{c}},\kappa_{\text{c}},\sigma,t),\\
    &(\delta J_z)^2 = \sum\limits_{n}\frac{(n-m_{\gamma})^2}{[(|m_{\gamma}-n|)!]^2}\left(\frac{\kappa_{\text{c}}\sigma_{\text{b}}}{2}\right)^{2|m_{\gamma}-n|}\mathcal{I}_{n,\lambda}(\Gamma,\omega_{\text{c}},\kappa_{\text{c}},\sigma,t).
\end{aligned}
\end{equation}

\bibliography{references}